\documentclass[prc,superscriptaddress,unsortedaddress,twocolumn,showpacs]{revtex4}
\usepackage{graphicx}
\usepackage{amsmath}
\usepackage{amssymb}
\usepackage{times}
\usepackage{comment}
\usepackage{bm}
\usepackage{braket}
\newcommand {\vecK}{\mathbf{K}}
\newcommand {\vecR}{\mathbf{R}}
\newcommand {\vecr}{\mathbf{r}}

\usepackage{color}
\usepackage{ulem}

\def\fun#1#2{\lower3.6pt\vbox{\baselineskip0pt\lineskip.9pt
\ialign{$\mathsurround=0pt#1\hfil##\hfil$\crcr#2\crcr\sim\crcr}}}
\begin{document}

\title{
Benchmarking theoretical formalisms for $(p,pn)$ reactions: the $^{15}$C($p,pn$)$^{14}$C case
}

\author{K. Yoshida}
\email[]{yoshidak@rcnp.osaka-u.ac.jp}
\affiliation{Research Center for Nuclear Physics (RCNP), Osaka
University, Ibaraki 567-0047, Japan}

\author{M. G\'omez-Ramos}
\affiliation{Departamento de FAMN, Universidad de Sevilla, Apartado 1065, 41080 Sevilla, Spain}

\author{K. Ogata}
\affiliation{Research Center for Nuclear Physics (RCNP), Osaka
University, Ibaraki 567-0047, Japan}

\author{A. M. Moro}
\affiliation{Departamento de FAMN, Universidad de Sevilla, Apartado 1065, 41080 Sevilla, Spain}

\date{\today}

\begin{abstract}
\begin{comment}
%We test the validity of the distorted-wave impulse approximation (DWIA)  and the transfer to the continuum model (TC) for 
$^{15}$C($p$,$pn$)$^{14}$C reaction calculation by comparing with 
Faddeev/Alt-Grassberger-Sandhas (FAGS) result.
It has shown that both DWIA and TC are in fairly good agreement with 
FAGS.
It has also confirmed through the investigation on DWIA results
that the energy dependence of distorted waves of the emitted 
nucleons gives only a few percent difference on knockout cross sections.
Therefore one can safely fix the energy parameters for the distorting
potentials of the emitted particles, which is the almost unavoidable
treatment in TC and FAGS formalism.
\end{comment}

\begin{description}
\item[Background] Proton-induced knockout reactions of the form $(p,pN)$ have experienced a renewed interest in recent years due to the possibility of performing these measurements with rare isotopes, using inverse kinematics. Several theoretical models are being used for the interpretation of these new data, such as the distorted-wave impulse approximation (DWIA), the transition amplitude formulation of the Faddeev equations due to Alt, Grassberger and Sandhas (FAGS) and, more recently, a coupled-channels method here referred to as transfer-to-the-continuum (TC).   

\item[Purpose] Our goal is to compare the momentum distributions calculated with the DWIA and TC models for the same reactions, using whenever possible the same inputs (e.g., distorting potential). A comparison with already published results for the FAGS formalism is performed as well.

\item[Method] We choose the $^{15}$C($p$,$pn$)$^{14}$C reaction at an incident energy of 420~MeV/u, which has been previously studied with the FAGS formalism. The knocked-out neutron is assumed to be in a $2s$ single-particle orbital. Longitudinal and transverse momentum distributions are calculated for different assumed separation energies.   

\item[Results]  For all cases considered, we find a very good agreement between DWIA and TC results. The energy dependence of the distorting optical potentials  is found to affect in a modest way the shape and magnitude of the momentum distributions.  Moreover, when relativistic kinematics corrections are omitted, our calculations reproduce remarkably well the FAGS result. 

\item[Conclusions] The results found in this work provide confidence on the consistency and accuracy of the DWIA and TC models for analyzing  momentum distributions for 
 $(p,pn)$  reactions at intermediate energies.

\end{description}
\end{abstract}

\pacs{24.10.Eq, 25.40.-h}
%24.10.Eq Coupled-channel and distorted-wave models
%25.40.-h Nucleon-induced reactions

\maketitle

\section{Introduction}
Thanks to the  development of radioactive isotope beam technology,
experiments on unstable nuclei 
in inverse kinematics have been made possible. Among them,  studies on 
single-particle structure and 
its evolution in nuclei away from the stability valley
is one of the main subjects of study in present day nuclear physics.
 Knockout reactions induced by intermediate energy protons have been one of the most successful tools for studying the  single-particle nature both of stable and unstable nuclei.
The distorted-wave impulse approximation (DWIA) 
is one of the reaction models which has been successfully
applied to the analysis of these reactions  \cite{Sam86,Cha77,Sam87,Jac66,Jac73,Kit85} (for a recent review, see Ref.~\cite{Wak17}). Most DWIA applications have been done for  exclusive measurements and under  {\it quasi-free} scattering conditions. It remains to assess  the accuracy of the method for more inclusive observables, such as total nucleon removal cross sections and momentum distributions of the residual heavy fragment. A recent step toward this goal is provided by the  eikonal  DWIA formalism  recently proposed in Ref.~\cite{Aum13}.

In recent years, the Alt-Grassberger-Sandhas  formulation of the Faddeev equations (FAGS) 
\cite{Fad61,Alt67}, which uses a momentum-space representation of the scattering transition amplitude, has been 
put forward as an alternative for the analysis of these kinds of processes 
\cite{Cre08,Cre09,Cre14,Cra16}.

Very recently, another reaction model, referred to as  
the transfer-to-the-continuum (TC) framework, has been 
developed and applied to ($p$,$pN$) reactions
\cite{Mor15,Mar17}. Since these three formalisms are being used to analyze experimental data, it is of timely importance to establish the consistency among them, and understand the limitations and range of validity in each case. 

Within the same scope, it has been shown in \cite{Cre08,Cre09} that one can recover the DWIA formalism using a truncated Faddeev multiple-scattering series. However, the DWIA so obtained differs in some aspects from the one commonly used in actual analyses of $(p,pN)$ data, since the latter usually involves additional approximations. 

It is therefore essential to make a comparison
between these models and, as a first step towards this goal,  
 in this paper
we make a benchmark comparison between DWIA, TC, and FAGS, for a given  ($p$,$pn$) reaction using, whenever possible, the same input ingredients in the calculations.

The content of the paper is as follows.
In Sec.~\ref{secformalism} the formulation of the DWIA and the TC
formalisms is given.
In Sec.~\ref{secresult} the longitudinal momentum distributions (LMDs) of the $^{15}$C($p$,$pn$)$^{14}$C reaction 
with DWIA an TC are compared, for different separation energies and studying the effect of the energy dependence of the distorting potentials for the emitted nucleons. A comparison with the FAGS transversal momentum distributions (TMDs) published in Ref.~\cite{Cra16} is also presented.
Finally, the summary is given in Sec.~\ref{secsum}.

\section{Formalism}
\label{secformalism}
We consider A($p$,$pn$)B knockout reaction in inverse kinematics.
Observables shown below with superscript A are evaluated in the
so-called A-rest frame.

\subsection{DWIA framework}
\label{subsecdwia}
In the DWIA formalism, the transition amplitude for a A($p$,$pn$)B reaction
is given by
\begin{align}
T_{\bm{K}_0\bm{K}_1\bm{K}_2}^{nljm}
&=
\Braket{
\chi_{1,\bm{K}_1}^{(-)}\chi_{2,\bm{K}_2}^{(-)}
|t_{pn}|
\chi_{0,\bm{K}_0}^{(+)}\varphi^{nljm}
},
\label{eqtrans}
\end{align}
where $\chi_0$, $\chi_1$, and $\chi_2$ are the scattering wave functions
of the $p$-A, $p$-B, and $n$-B systems, respectively,
$\varphi^{nljm}$ is the single-particle wave function with 
$n$, $l$, $j$, and $m$ being the principal quantum number, 
the orbital angular momentum, the total angular momentum, 
and its third component of $n$ bound in A, respectively.
The transition interaction $t_{pn}$ is the effective interaction 
between $p$-$n$ pair which reproduces $p$-$n$ binary scattering.

By applying the so-called factorization approximation,
which has been reconfirmed to be valid in Ref.~\cite{Yos16},
Eq.~(\ref{eqtrans}) is reduced to 
\begin{align}
T_{\bm{K}_0\bm{K}_1\bm{K}_2}^{nljm}
&\approx
\Braket{
\bm{\kappa}'
|t_{pn}|
\bm{\kappa}
}
\int d\bm{R}\,
F(\bm{R})\varphi^{nljm}(\bm{R}),
\label{eqfact}
\end{align}
where $F(\bm{R})$ is defined by
\begin{align}
F(\bm{R})
\equiv
\chi_{1,\bm{K}_1}^{*(-)}(\bm{R})\,
\chi_{2,\bm{K}_2}^{*(-)}(\bm{R})\,
\chi_{0,\bm{K}_0}^{(+)}(\bm{R})\,
e^{-i\bm{K}_0\cdot\bm{R}/A}.
\end{align}
The initial and final relative momenta of the $p$-$n$ system are defined by
\begin{align}
\bm{\kappa}
&\equiv
\left(\alpha_0\bm{K}_0-\bm{K}_n\right)/2,\\
\bm{\kappa}'
&\equiv
\left(\bm{K}_1-\bm{K}_2\right)/2,
\end{align}
with $\alpha_0=(A+1)/A$.
The momentum of $n$ in the initial state $\bm{K}_n$ is evaluated
from asymptotic momenta 
by assuming the momentum conservation in the $p$-$n$ system:
\begin{align}
\bm{K}_n=\bm{K}_1+\bm{K}_2-\alpha_0\bm{K}_0.
\end{align}
In the present study on-the-energy-shell (on-shell) approximation is
adopted in taking the squared modulus of Eq.~(\ref{eqfact}):
\begin{align}
\frac{\mu_{pn}^2}{(2\pi\hbar^2)^2}
\lvert
\Braket{
\bm{\kappa}'
|t_{pn}|
\bm{\kappa}
}
\rvert ^2
\approx
\frac{d\sigma_{pn}}{d\Omega_{pn}}
\left(E_{pn},\theta_{pn}\right),
\end{align}
where $\mu_{pn}$ is the reduced mass of the $p$-$n$ system,
$\theta_{pn}$ is the angle between $\bm{\kappa}'$ and $\bm{\kappa}$,
and the $p$-$n$ scattering energy is given by
\begin{align}
E_{pn}
&=
\frac{\hbar^2(\kappa^2 +\kappa'^2)/2}{2\mu_{pn}}
.
\end{align}

In the present DWIA, the momentum distribution (MD) is given by
\begin{align}
\frac{d\sigma}{d\bm{K}_\mathrm{B}^\mathrm{A}}
=
&C_0 \int d\bm{K}_1^\mathrm{A}d\bm{K}_2^\mathrm{A}\eta_\textrm{M\o l}^\mathrm{A}
\delta(E_f^\mathrm{A}-E_i^\mathrm{A})
\delta(\bm{K}_f^\mathrm{A}-\bm{K}_i^\mathrm{A}) \nonumber \\
&\times 
\frac{d\sigma_{pn}}{d\Omega_{pn}}(E_{pn},\theta_{pn})
\sum_{m}(2\pi)^2
\lvert
\bar{T}_{\bm{K}_0\bm{K}_1\bm{K}_2}^{nljm}
\rvert ^2,
\end{align}
where
\begin{align}
C_0
&\equiv
\frac{E_0^\mathrm{A}}{(\hbar c)^2 K_0^\mathrm{A}}
\frac{1}{(2l+1)}\frac{\hbar^4}{(2\pi)^3 \mu_{pn}^2}, \\
\eta_\textrm{M\o l}^\mathrm{A}
&\equiv
  \frac{E_1 E_2 E_\mathrm{B}}{E_1^\mathrm{A} E_2^\mathrm{A} E_\mathrm{B}^\mathrm{A}},
\end{align}
and the reduced transition amplitude is given by
\begin{align}
\bar{T}_{\bm{K}_0\bm{K}_1\bm{K}_2}^{nljm}
&=
\int d\bm{R}\,F(\bm{R})\varphi^{nljm}(\bm{R}).
\end{align}
Longitudinal and transverse MD are obtained from MD as follows:
\begin{align}
\frac{d\sigma}{dK_{\mathrm{B}z}^\mathrm{A}}
&=
2\pi\int dK_{\mathrm{B}b}^\mathrm{A} K_{\mathrm{B}b}^\mathrm{A}
\frac{d\sigma}{d\bm{K}_\mathrm{B}^\mathrm{A}}, \\
\frac{d\sigma}{dK_{\mathrm{B}x}^\mathrm{A}}
&=
\int dK_{\mathrm{B}y}^\mathrm{A}dK_{\mathrm{B}z}^\mathrm{A}
\frac{d\sigma}{d\bm{K}_\mathrm{B}^\mathrm{A}}.
\end{align}

%-------------------------------------------
\subsection{Transfer to the continuum model}
%---------------------------------------------
The transfer to the continuum formalism is based on the prior representation of the transition matrix for the A($p$,$pn$)B process:
\begin{equation}
T_{if}^{nljm}=
\Braket{
\Psi_f^{3b(-)}
|V_{pn}+V_{p\mathrm{B}}-U_{p\mathrm{A}}|
\chi_{0,\bm{K}_0}^{(+)}\varphi^{nlj m}
},
\end{equation}
where $\varphi^{nlj m}$ and $\chi_{0,\bm{K}_0}^{(+)}$ are defined as above, $V_{pn}$ and $V_{p\mathrm{B}}$ are the corresponding binary potentials, $U_{p\mathrm{A}}$ is the optical potential used to generate the distorted wave $\chi_{0,\bm{K}_0}^{(+)}$, and $\Psi_f^{3b(-)}$ is the final state wavefunction, which is formally treated as a three-body wavefunction, under the approximation that the state of B is not modified during the reaction.

In order to perform the calculation, $\Psi_f^{3b(-)}$ is expanded in terms of $p+n$ eigenstates, that is,
\begin{equation}
\label{eq:Phi3b}
\Psi_f^{3b(-)}(\vecr,\vecR)=\sum_{j'\pi}\int \mathrm{d}k \phi^{j'\pi}(k,\vecr)\chi^{j'\pi}(\vecK, \vecR),
\end{equation}
where $k$ is the relative wave number of the $p+n$ pair and $K$ is the wave number for the relative motion between B and the $p+n$ pair and is related to $k$ through energy conservation. $\phi^{j'\pi}(k,\vecr)$ are the eigenstates for the $p+n$ Hamiltonian with the interaction $V_{pn}$ and wave number $k$, total angular momentum $j$ and parity $\pi$ while $\chi^{j'\pi}(\vecK, \vecR)$ describes the motion of the $p+n$ pair with respect to B for a wave number $\mathrm{K}$, with the $p+n$ pair having total momentum $j$ and parity $\pi$. Note that the expansion  (\ref{eq:Phi3b}) contains also the term with the bound deuteron. This term is omitted here for brevity. 

The $k$ continuum is discretized using a binning procedure in a similar way to continuum-discretized coupled-channel calculations (CDCC),
\begin{equation}
\Psi_f^{3b(-)}(\vecr,\vecR) \approx\sum_{Nj'\pi} \phi_N^{j'\pi}(k_N,\vecr)\chi_N^{j'\pi}(\vecK_N,\vecR),
\label{eq3bwf}
\end{equation}
where $k_N$ is an average momentum of the bin, and $\phi_N$ are the bin wave functions. As such, the transition matrix results in
\begin{align}
T_{if} &\approx \sum_{Nj'\pi}\langle
\phi_N^{j'\pi} \chi_N^{j'\pi}
|V_{pn}  %\nonumber \\ 
 + V_{p\mathrm{B}}-U_{p\mathrm{A}}|
\chi_{0,\bm{K}_0}^{(+)}\varphi^{nlj m}
\rangle.
\end{align}

In order to make a more meaningful comparison with the DWIA calculations, the terms $V_{p\mathrm{B}}-U_{p\mathrm{A}}$ have been ignored, in what is called the \textit{no-remnant} approximation,
\begin{align}
T_{if} & \approx % \\\nonumber
\sum_{Nj'\pi}\Braket{
\phi_N^{j'\pi} \chi_N^{j'\pi}
|V_{pn}|
\chi_{0,\bm{K}_0}^{(+)}\varphi^{nlj m}
}.
\label{eqnoremnant}
\end{align}

This transition amplitude is computed employing a calculation akin to a coupled-channel Born approximation (CCBA), from which the angular differential cross section to each of the bin states can be computed.
A double differential cross section on the outgoing angle of B and the internal energy of the $p+n$ pair can be obtained from the angular differential cross section to each of the bins through
\begin{equation}
\left . \dfrac{\mathrm{d}^2\sigma_{j'\pi}}{\mathrm{d}\epsilon_{pn}\mathrm{d}\Omega_B}  \right|_{\epsilon_{pn} \in \Delta_N}
\approx \dfrac{1}{\Delta_N}\dfrac{\mathrm{d}\sigma_{N,j'\pi}}{\mathrm{d}\Omega_B},
\end{equation}
where $\Delta_N$ is the energy width of the  bin $\{N,j',\pi\}$. Through energy conservation and the proper Jacobians, the longitudinal and transverse momentum distributions of B can be obtained from this double differential cross section.	

From the practical point of view, an appealing feature of the TC method is that the sum in Eqs.~(\ref{eq3bwf})--(\ref{eqnoremnant}) converges with a few values of $j'$ (typically $j<4$ at the intermediate energies considered here). A limitation is however that the interactions $p$+B and $n$+B are assumed to be energy independent. The accuracy of this will be investigated in the next section by comparing with the DWIA calculations.

The transfer to the continuum calculations have been performed using a modified version of the code {\sc fresco} \citep{Tho88}. Further details can be found in \citep{Mor15}.

%--------------------------------
\section{Results and discussion \label{secresult}}
%--------------------------------
In this section, we compare the calculations with the TC and DWIA methods described in the preceding section. We consider the reaction $^{15}$C($p$,$pn$)$^{14}$C, calculating the knockout of a neutron in a $2s$ single-particle orbital and for three different separation energies: $S_n=1.22$~MeV (i.e.\ the physical value), 5 MeV~and 18~MeV.   

\subsection{Numerical inputs}
\label{subsecinput}
The single-particle wave function of the struck neutron, $\varphi^{nljm}$, is obtained for a Woods-Saxon  central potential 
$V(R)=V_0/\left(
1+\mathrm{exp}\left[(R-r_0 B^{1/3})/a_0\right]
\right)
$
with $r_0=1.25$~fm and $a_0=0.65$~fm.
The depth parameter $V_0$ is adjusted so as to give neutron
separation energies $S_n=1.22$~MeV, $5$~MeV and $18$~MeV.

For the nucleon-nucleon interaction, we employ the Reid93 potential \cite{Reid93}, a generalized version of the pioneering Reid soft core potential \cite{Reid68}, developed by the Nijmegen group. This potential contains central, spin-orbit and tensor components, and reproduces accurately the proton-proton and proton-neutron phase-shifts up to an energy of 350~MeV ($\chi^2/N_\mathrm{data}=1.03$). 

As for the distorting potential of $p$-A, $p$-B, and $n$-B systems, we use the 
EDAD2 parameter set of the Dirac phenomenology \cite{Ham90}. In the comparison with the FAGS calculations, the global optical potential parameters of Koning and Delaroche \cite{Kon03} will be also considered. 

%In the present results of DWIA, TC, and FAGS, different  $p$-$n$ interactions are employed, i.e.  Franey-Love effective interaction \cite{Lov81},  Reid93  \cite{Reid93}, and realistic CD-Bonn potential \cite{Mac01}, respectively.

%------------------------------------------
\subsection{Comparison between TC and DWIA}
\begin{figure}[htbp]
\centering
\includegraphics[width=0.40\textwidth]{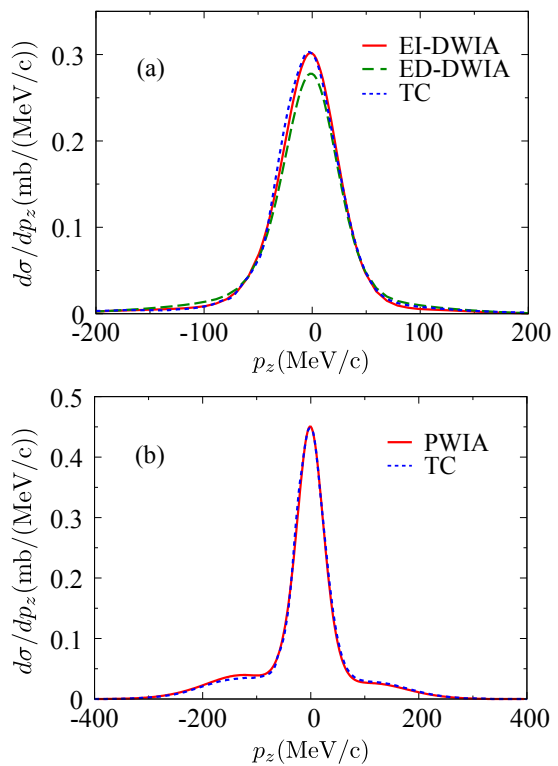}
\caption{
(Color online)
(a)
Longitudinal momentum distribution of the $^{15}$C($p$,$pn$)$^{14}$C
reaction at 420~MeV/u. The struck neutron is assumed to be in a $2s$ orbital with a separation energy of 1.22~MeV. The solid, dashed and dotted lines show the results of DWIA (energy-independent potentials), DWIA (energy-dependent potentials)  and TC, respectively.
(b) Same as (a) but with all distorting potentials  switched off. 
}
\label{figDWIA_TC_1.22}
\end{figure}

%-------------------------------------------
In Fig.~\ref{figDWIA_TC_1.22}(a) we compare the longitudinal momentum  distribution calculated with the DWIA and TC reaction frameworks. 
A separation energy of  $S_n=1.22$~MeV is assumed for the removed neutron in $^{15}$C. For the nucleon-nucleus distorting potentials we use the Dirac phenomenology \cite{Ham90}. For the incident proton, this potential was evaluated at 420~MeV, whereas for the outgoing nucleus the potential was evaluated at 210~MeV. 
It is seen that the TC and DWIA results are in excellent agreement, both in shape and magnitude.  From this comparison, we conclude that these reaction formalisms yield fully consistent results at this energy.

In Fig.~\ref{figDWIA_TC_1.22}(b) the same comparison with (a) but
without all distorting potentials are shown.	
It should be noted that the agreement in the plane wave limit
is worth investigating because a difference between DWIA and TC
may appear, 
since the distortion effects suppress the tail region of MD,
as shown in Figs.~\ref{figDWIA_TC_1.22}(a) and (b).
As a result, the perfect agreement is found in the plane wave limit
as well.

%--------------------------------------------------------
%\subsection{Energy dependence of distorting potentials}
%--------------------------------------------------------
%In the preceding calculations, the optical potentials for the outgoing nucleons was fixed to half of  the incident energy. 
Since the outgoing  nucleons are expected to emerge  with a broad range of energies, using optical potentials fixed at half of  the incident energy may not be a good approximation.
In DWIA, this effect can be readily taken into account by evaluating the outgoing distorted potentials at the energy given by their asymptotic momenta. To assess  the importance of this effect, in Fig.~\ref{figDWIA_TC_1.22}(a) we show also  the DWIA calculation including  this energy dependence (dashed line).   
%The DWIA method permits however to take into account the energy dependence of these potentials so in this section we study the importance of this effect by comparing DWIA calculations with or without this energy dependence. This is shown in  Fig.~\ref{figDWIA_TC_1.22}, where we show such calculations  effect of energy dependence  of distorting potential on the longitudinal momentum distribution of  $^{15}$C($p$,$pn$)$^{14}$C reaction. Solid, dashed, and dotted lines correspond to energy-independent(EI)-DWIA, energy-dependent(ED)-DWIA, and TC, respectively.

\begin{comment}
\begin{figure}[htbp]
\centering
\includegraphics[width=0.40\textwidth]{122MeV_revised.eps}
%\includegraphics[width=0.40\textwidth]{PWIA_200MeV.eps}
\caption{
(Color online)
(a)
Longitudinal momentum distribution of
$^{15}$C($p$,$pn$)$^{14}$C reaction at 420~MeV with 
$S_n=1.22$~MeV.
The solid, dashed, and dotted line shows 
energy-independent(EI)-DWIA, energy-dependent(ED)-DWIA, 
and TC result, respectively.
(b)
Same as (a) but all distorting potentials are switched off
in PWIA (solid) and TC (dashed).
}
\label{figDWIA_TC_1.22}
\end{figure}
\end{comment}

%The EDAD2 \cite{Ham90} optical potential is employed as a distorting potential, and the scattering energy parameter for distorting potentials of  emitted $p$ and $n$ is fixed at 200~MeV in EI-DWIA and TC calculations, while the energy dependence is treated properly  in EI-DWIA.
 One can see  that, by taking the 
energy dependence of distorted waves into account, the 
LMD is reduced by 8.0\% at the peak in the DWIA calculation so, at least for this system and incident energy,  
the energy dependence produces a minor, although non-negligible, effect.
%It is found that the ED-DWIA agrees better than the energy-independent one, 
%however, this agreement can be concluded to be accidental by
%looking at the results with plane waves shown in Fig.~\ref{figDWIA_TC_1.22} (b).
%Since the difference between PWIA and TC in the figure essentially originated
%from the different treatment of $p$-$n$ interactions,
%and this difference should exist also in the results with distorted waves.
%Therefore, the difference coming from the $p$-$n$ interactions is somewhat
%canceled out by considering the energy dependence of distorting potentials
%in Fig.~\ref{figDWIA_TC_1.22} (a).

%--------------------------------------
\subsection{Binding energy dependence}
\begin{figure}[htbp]
\centering
\includegraphics[width=0.40\textwidth]{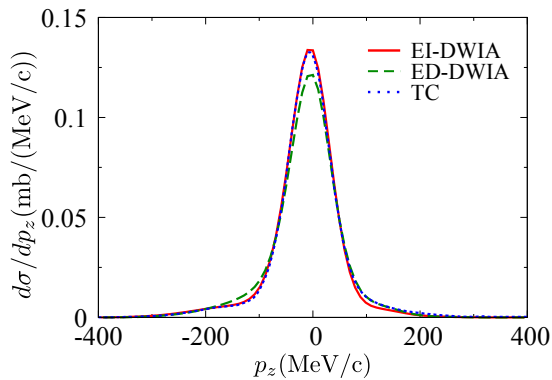}
\caption{
(Color online)
Same as in Fig.~\ref{figDWIA_TC_1.22}(a)
but for $S_n=5$~MeV.
}
\label{fig5MeV}
\end{figure}

\begin{figure}[htbp]
\centering
\includegraphics[width=0.40\textwidth]{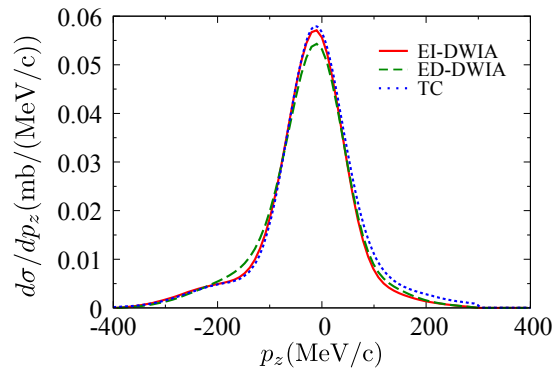}
\caption{
(Color online)
Same as in Fig.~\ref{figDWIA_TC_1.22}(a)
but for $S_n=18$~MeV.
}
\label{fig18MeV}
\end{figure}

%---------------------------------------
In this section we continue the benchmark test of
DWIA and TC changing by the neutron separation energy artificially.
In Figs.~\ref{fig5MeV} and \ref{fig18MeV}
LMD of $^{15}$C($p$,$pn$)$^{14}$C reaction with 
$S_n=5$~MeV and $18$~MeV are shown, respectively.

It is found that EI-DWIA, ED-DWIA, and TC also agree well 
in both $S_n=$ 5~MeV and 18~MeV cases and at the same level as in the 
$S_n=$ 1.22 MeV case. The LMD is reduced by 9.3\% 
(4.9\%) at the peak when 
$S_n=$ 5~MeV (18~MeV) by taking
the energy dependence of the optical potential parameters
for the emitted $p$ and $n$. 
It is also found that 
the asymmetric shape of LMD due to the asymmetry of the phase space,
which is discussed in Ref.~\cite{Oga15}, 
is gradually developed as $S_n$ increases through 
Figs.~\ref{figDWIA_TC_1.22}--\ref{fig18MeV}.

%---------------------------------
\subsection{Comparison with FAGS}
\begin{figure}[htbp]
\centering
\includegraphics[width=0.40\textwidth]{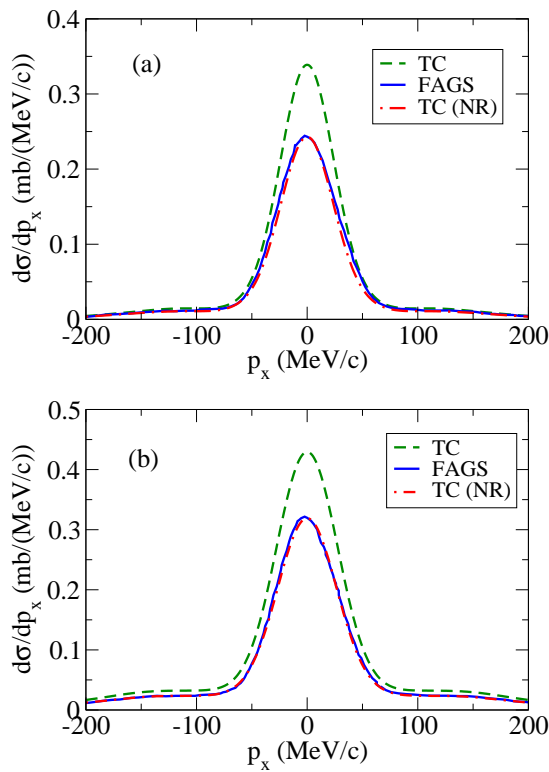}
\caption{
(Color online)
(a)
Transversal momentum distribution of $^{15}$C($p$,$pn$)$^{14}$C
reaction at 420~MeV/u. The solid line is the  FAGS result taken from Fig.~4 of Ref.~\cite{Cra16}. The dashed and dot-dashed lines are the TC calculations with and without relativistic corrections, respectively.
(b)
Same as (a) but with all distorting potentials switched off.}
\label{figcravo}
\end{figure}

%---------------------------------
Finally, we compare our calculations with the more sophisticated Faddeev-AGS (FAGS) framework. 
%In this subsection, we compare our results with those obtained with the more sophisticated Faddeev-AGS (FAGS) framework. 
This in presented in Fig.~\ref{figcravo}, where we show the transverse momentum distribution for the $^{15}$C($p$,$pn$)$^{14}$C reaction at 420~MeV/u with $S_n=1.22$~MeV. As in previous cases, the upper and bottom panels correspond to the full calculations and the calculations assuming plane waves for the incoming and outgoing nucleons. In each panel, the solid line is the FAGS calculation,  taken from Fig.~4 of Ref.~\cite{Cra16}. This calculation was performed  with the Koning-Delaroche nucleon-nucleus potential, evaluated at 200 MeV, and assuming non-relativistic kinematics. The dot-dashed line is the TC calculation using the same optical potential without any relativistic corrections for consistency. The agreement between these two calculations is excellent. 
It is to be noted that the FAGS calculation employs the CD-Bonn $NN$ potential~\cite{Mac01}, whereas our TC implementation uses the Reid93 potential. These two $NN$ potentials yield essentially the same on-shell observables up to 350~MeV, so we believe that, despite this different choice, the comparison is meaningful.  

To highlight the importance of relativistic effects, we depict also in this figure the TC calculation including relativistic kinematics corrections (dashed line). It is seen that these corrections have a small effect on the shape of the momentum distribution, but they increase significantly its magnitude by about 30\%. Consequently, the inclusion  of these relativistic effects will be relevant for the extraction of reliable spectroscopic factors  from the analysis of ($p$,$pn$) data. 

The same calculations shown in Fig~\ref{figcravo} (a) but switching off the distorting potential of the incoming and outgoing nucleons are shown in Fig.~\ref{figcravo} (b) to see clearly the difference arising from a different choice of $NN$ potentials.
One can see that the good agreement between TC and FAGS remains in this case.

\begin{comment}
Both DWIA and TC \textcolor{red}{reproduce} fairly well \sout{reproduce} \textcolor{red}{the} FAGS distribution;
\textcolor{red}{the} integrated cross section of DWIA (TC) deviates  \sout{\bf 8.6\% (2.1\%)} from
that of FAGS.
The K\"{o}ning-Delaroche optical potential parameters~\cite{Kon03} 
at 200~MeV are adopted for all nucleon scattering states
as in the same way as Ref.~\cite{Cra16}.

It should be noted that 
those difference in plane wave result should be 
mainly originated from the different treatment in 
$p$-$n$ interactions.
PWIA, TC and FAGS employs Franey-Love effective interaction,
Reid soft core potential, and realistic CD-Bonn potential,
respectively.

\end{comment}

%------------------------
\section{Summary}
\label{secsum}
%------------------------
Transverse and longitudinal momentum distribution 
of the residual $^{14}$C  nucleus produced in the $^{15}$C($p$,$pn$)$^{14}$C knockout reaction at an incident energy of 420~MeV/u have been computed  and compared using   
different reaction frameworks, namely, the distorted-wave impulse approximation (DWIA), the transfer-to-the-continuum (TC) method, and the Faddeev-AGS (FAGS) formalism. 

The longitudinal momentum distributions evaluated with TC and EI-DWIA are found to be in excellent agreement both in the shape and magnitude.
The agreement remains for increasing separation energies of the removed neutron,
giving only 0.3\%, 0.8\%, 1.4\% difference at the peak 
when $S_n=$ 1.22~MeV, 5~MeV, 18~MeV, respectively,
corroborating the consistency of the two methods for weakly-bound and tightly bound systems. 
We found that the energy dependence
of the optical potentials for emitted nucleons,
which are not taken into account in TC, gives
a minor, although non-negligible effect on knockout cross section.

The TC calculation, omitting relativistic kinematics corrections, is also found to reproduce remarkably well the FAGS calculation reported for this reaction. However, the inclusion of relativistic corrections increases the TC result by $\sim$30\%, which highlights the relevance of these effects  for the extraction of spectroscopic information from absolute $(p,pN)$ cross sections. 

%Longitudinal momentum distributions of DWIA and TC  also give a good agreement, and
%it was confirmed that the 
%It should be noted again that present calculations have  been performed with the same $p$-A, $p$-B, $n$-B distorting and binding potentials but different $p$-$n$ interactions.

From this study, we conclude that the DWIA and TC methods can be reliably used to analyze the momentum distributions for  $(p,pn)$ cross sections, which are currently being measured by several experimental campaigns. Extensions of the present benchmark to other situations, such as the $(p,2p)$ case  or the removal from non $s$-wave nucleons, are in progress and will be published elsewhere.  

%\section*{ACKNOWLEDGMENTS}
\begin{acknowledgments}
This work has received funding from the Spanish Ministerio de
Econom\'{i}a y Competitividad under Project No. FIS2014-53448-C2-
1-P and by the European Union Horizon 2020 research and innovation program under Grant Agreement 
No.\ 654002. M.G.-R. acknowledges support from the Spanish Ministerio de Educaci\'on,
Cultura y Deporte, Research Grant No. FPU13/04109.
A part of the computation was carried out with the computer
facilities at the Research Center for Nuclear Physics, Osaka
University. This work was supported in part by Grants-in-Aid
of the Japan Society for the Promotion of Science (Grants No. JP16K05352
and No. JP15J01392), and RCNP Young Foreign
Scientist Promotion Program.
\end{acknowledgments}

%%--------------------------------------------------------------------%%
%%                           References                               %%
%%--------------------------------------------------------------------%%


\begin{thebibliography}{00}
\bibitem{Sam86}
C.~Samanta, N.~S.~Chant, P.~G.~Roos, A.~Nadasen, J.~Wesick, and A.~A.~Cowley, 
Phys. Rev. C \textbf{34}, 1610 (1986).

\bibitem{Cha77}
N.~S.~Chant and P.~G.~Roos, 
Phys. Rev. C \textbf{15}, 57 (1977).

\bibitem{Sam87}
C.~Samanta, N.~S.~Chant, P.~G.~Roos, A.~Nadasen, and A.~A.~Cowley, 
Phys. Rev. C \textbf{35}, 333 (1987).

\bibitem{Jac66}
G.~Jacob and Th.~A.~J.~Maris, 
Rev. Mod. Phys. \textbf{38}, 121 (1966).

\bibitem{Jac73}
G.~Jacob and Th.~A.~J.~Maris, 
Rev. Mod. Phys. \textbf{45}, 6 (1973).

\bibitem{Kit85}
P.~Kitching, W.~J.~McDonald, Th.~A.~J.~Maris, and C.~A.~Z.~Vasconcellos, 
Adv. Nucl. Phys. \textbf{15}, 43 (1985).

\bibitem{Wak17}
T.~Wakasa, K.~Ogata, and T.~Noro,
Prog. Part. Nucl. Phys. \textbf{96}, 32 (2017).

\bibitem{Aum13}
T. Aumann, C. A. Bertulani, and J. Ryckebusch
Phys. Rev. C \textbf{88}, 064610 (2013).

\bibitem{Fad61}
L.~D.~Faddeev, 
Zh. Eksp. Teor. Fiz. \textbf{39}, 1459 (1960)
[Sov. Phys. JETP \textbf{12}, 1014 (1961)].

\bibitem{Alt67}
E.~O.~Alt, P.~Grassberger, and W.~Sandhas, 
Nucl. Phys. B \textbf{2}, 167 (1967).

\bibitem{Cre08}
R.~Crespo, A.~Deltuva, E.~Cravo, M.~Rodr\'{i}guez-Gallardo, and A.~C.~Fonseca,
Phys. Rev. C \textbf{77}, 024601 (2008).

\bibitem{Cre09}
R.~Crespo, A.~Deltuva, M.~Rodr\'{i}guez-Gallardo, E.~Cravo, and A.~C.~Fonseca,
Phys. Rev. C \textbf{79}, 014609 (2009).

\bibitem{Cre14}
R.~Crespo, A.~Deltuva, and E.~Cravo,
Phys. Rev. C \textbf{90}, 044606 (2014).

\bibitem{Cra16}
E.~Cravo, R.~Crespo, and A.~Deltuva,
Phys. Rev. C \textbf{93}, 054612 (2016).

\bibitem{Mor15}
A.~M.~Moro,
Phys. Rev. C \textbf{92}, 044605 (2015).

\bibitem{Mar17}
M.~G\'{o}mez-Ramos J.~Casal, and A.~M.~Moro,
Phys. Lett. B \textbf{772} 115 (2017).

\bibitem{Reid93}
V. G. J. Stoks, R. A. M.  Klomp,  C. P. F. Terheggen, and J. J.  de Swart,
Phys. Rev. C\textbf{49}, 2950 (1994), doi:10.1103/PhysRevC.49.2950

\bibitem{Reid68}
R. V. {Reid, Jr.}, Ann. Phys. (NY) \textbf{50}, 411 (1968). 

\bibitem{Yos16}
K.~Yoshida, K.~Minomo, and K.~Ogata,
Phys. Rev. C \textbf{94}, 044604 (2016).

\bibitem{Kon03}
A.~J.~Koning and J.~P.~Delaroche, 
Nucl. Phys. A \textbf{713}, 231 (2003).

\bibitem{Ham90}
S.~Hama, B.~C.~Clark, E.~D.~Cooper, H.~S.~Sherif, and R.~L.~Mercer,
Phys. Rev. C \textbf{41}, 2737 (1990);
E.~D.~Cooper, S.~Hama, B.~C.~Clark, and R.~L.~Mercer,
\textit{ibid}. \textbf{47}, 297 (1993).

\bibitem{Mac01}
R.~Machleidt, Phys. Rev. C \textbf{63}, 024001 (2001).

\bibitem{Oga15}
K.~Ogata, K.~Yoshida and K.~Minomo,
Phys. Rev. C \textbf{92}, 034616 (2015).

\bibitem{Tho88}
I.J.~Thompson,
Computer Physics Reports \textbf{7}(4), 167 (1988).
\end{thebibliography}
\end{document}